\documentclass[12pt]{article}
\begin{document}
\title{BOSONS, FERMIONS AND BRANES}
\author{B.G. Sidharth\\
International Institute for Applicable Mathematics \& Information Sciences\\
Hyderabad (India) \& Udine (Italy)\\
B.M. Birla Science Centre, Adarsh Nagar, Hyderabad - 500 063 (India)}
\date{}
\maketitle
\begin{abstract}
Starting from considerations of Bosons at the real life Compton
scale we go on to a description of Fermions, specifically the Dirac
equation in terms of an underlying noncommutative geometry described
by the Dirac $\gamma$ matrices and generalize this to branes in an
underlying C-space (with Clifford Algebra).
\end{abstract}
\section{Bosonic Strings}
T. Regg's work of the 1950s \cite{reg,roma,tassie} as is well known,
showed that the resonances could be fitted by a straight line plot
in the $(J,M^2)$ plane, where $J$ denotes the angular momentum and
$M$ the mass of the resonances:
\begin{equation}
J \propto M^2,\label{ex}
\end{equation}
Equation (\ref{ex}) suggested that not only did resonances have
angular momentum, but they also resembled extended objects rather
than be point like. The problem is that if there is a finite extension for the electron
then forces on different parts of the electron would exhibit a time lag, requiring the
so called Poincare stresses for stability \cite{rohr,barut,feynman}. Interestingly
the electron was also modelled as a membrane \cite{bgsdirac}.\\
In 1968, G. Veneziano came up with a unified description of the Regge resonances (\ref{ex}) and other scattering processes. Veneziano considered the collision and scattering process as a black box and pointed out that there were in essence, two scattering channels, $s$ and $t$ channels. These, he argued gave a dual description of the same process \cite{ven,venezia}.\\
In an $s$ channel, particles A and B collide, form a resonance which quickly disintegrates into particles C and D. On the other hand in a $t$ channel scattering  particles A and B approach each other, and interact via the exchange of a particle $q$. The result of the interaction is that particles C and D emerge. If we now enclose the resonance and the exchange particle $q$ in an imaginary black box, it will be seen that the $s$ and $t$ channels describe the same input and the same output: They are essentially the same.\\
Let us now come to the inter-quark potential \cite{lee,smolin}.
There are two interesting features of this potential. The first is
that of confinement, which is given by a potential term like
$$V (r) \approx \sigma r, \quad r \to \infty ,$$
where $\sigma$ is a constant. This describes the large distance behavior between two quarks. The confining potential ensures that quarks do not break out of their bound state, which means that effectively free quarks cannot be observed.\\
The second interesting feature is asymptotic freedom. This is realized by a Coulumbic potential
$$V_c (r) \approx - \frac{\propto (r)}{r} (\mbox{small}\, r)$$
$$\mbox{where}\, \propto (r) \sim \frac{1}{ln(1/\lambda^2r^2)}$$
The constant $\sigma$ is called the string tension, because there are string models which yield $V(r)$. This is because, at large distances the inter-quark field is string like with the energy content per unit length becoming constant. Use of the angular momentum - mass relation indicates that $\sigma \sim (400 MeV)^2$.\\
Such considerations lead to strings which are governed by the equation \cite{walter,fog,bgskluwer,st}
\begin{equation}
\rho \ddot {y} - T y'' = 0,\label{e1}
\end{equation}
\begin{equation}
\omega = \frac{\pi}{2l} \sqrt{\frac{T}{\rho}},\label{e2}
\end{equation}
\begin{equation}
T = \frac{mc^2}{l}; \quad \rho = \frac{m}{l},\label{e3}
\end{equation}
\begin{equation}
\sqrt{T/\rho} = c,\label{e4}
\end{equation}
$T$ being the tension of the string, $l$ its length and $\rho$ the line density and $\omega$ in (\ref{e2}) the frequency. The identification (\ref{e2}),(\ref{e3}) gives (\ref{e4}), where $c$ is the velocity of light, and (\ref{e1}) then goes over to the usual d'Alembertian or massless Klein-Gordon equation.\\
Further, if the above string is quantized canonically, we get
\begin{equation}
\langle \Delta x^2 \rangle \sim l^2.\label{e5}
\end{equation}
Thus the string can be considered as an infinite collection of
harmonic oscillators \cite{fog}. Using equations (\ref{e2}) and
(\ref{e3}) the extension $l$ turns out to be of the order of the Compton wavelength in (\ref{e5}), a circumstance that was called one of the miracles of the string theory by Veneziano \cite{ven}.\\
We have described a ``Bosonic String'', in the sense that there is
no room for the Quantum Mechanical spin. This can be achieved by
giving a rotation to the relativistic quantized string as was done
by Ramond \cite{uof,ramond}. In this case we recover (\ref{ex}) of
the Regge trajectories. The particle is now an extended object, at
the Compton scale, rotating with the velocity of light. Furthermore
in superstring theory there is an additional term $a_0$, viz.,
\begin{equation}
J \leq (2\pi T)^{-1} M^2 + a_0 \hbar, \, \mbox{with}\, a_0 = +1 (+2) \, \mbox{for \, the \, open\, (closed)\, string.}\label{e6}
\end{equation}
The term $a_0$ in (\ref{e6}) comes from the Zero Point Energy. Usual gauge bosons are
described by $a_0 = 1$ and gravitons by $a_0 = 2$.\\
String theory has to deal with extra dimensions which reduce to the usual four
dimensions of physical spacetime when we invoke the Kaluza-Klein approach at the
Planck Scale \cite{kaluza}.\\
All these considerations have been leading to more and more complex models, the
latest version being the so called M-Theory. In this latest theory supersymmetry
is broken so that the supersymmetric partner particles do not have the same mass as the
known particle. Particles can now be described as soliton like branes, resembling the
earlier Dirac membrane. M-Theory also gives an interface with Black Hole Physics.
The advantage of Supersymmetry is that a framework is now available for the unification of
all the interactions including gravitation. It may be mentioned that under a SUSY
transformation, the laws of physics are the same for all observers, which is the
case in General Relativity (Gravitation) also. Under SUSY there can be a maximum
of eleven dimensions, the extra dimensions being curled up as in Kaluza-Klein theories.
In this case there can only be an integral number of waves around the circle, giving rise
to particles with quantized energy. However for observers in the other four dimensions, it
would be quantized charges, not energies. The unit of charge would depend on the radius of
the circle, the Planck radius yielding the value $e$. This is the root of the unification
of electromagnetism and gravitation in these theories.\\
There is still no contact with experiment. It also appears that these theories lead to
an unacceptably high cosmological constant and finally to a landscape of some $10^{500}$
universes so that the anthropic principle needs to be invoked to identify our universe.\\
The non-verifiability of the above considerations and the fact that
the Planck scale $\sim 10^{20}GeV$ is also beyond forseeable
attainment in collidors has lead to much recent criticism.
\section{Fuzzy Spacetime and Fermions}
Can we take an alternative route to use Bosonic Strings which are at
the real world Compton scale to obtain a description of Fermions
without going to the Planck scale? We saw above that Bosonic
particles could be described as extended objects at the Compton
scale. Given a minimum spacetime scale $a$, it was shown by Snyder
that, the following Lorentz invariant relations hold,
$$[x,y] = (\imath a^2/\hbar )L_z, [t,x] = (\imath a^2/\hbar c)M_x, etc.$$
\begin{equation}
[x,p_x] = \imath \hbar [1 + (a/\hbar )^2 p^2_x]; \cdots\label{e3z}
\end{equation}
If $a^2$ in (\ref{e3z}) is neglected, then we get back the usual canonical commutation
relations of Quantum Mechanics. This limit to an established theory is another attractive
feature of (\ref{e3z}).\\
However if order of $a^2$ is retained then the first of equations
(\ref{e3z}) characterize a completely different spacetime geometry,
one in which the coordinates do not commute. This is a
noncommutative geometry and indicates that spacetime within the
scale defined by $a$ is ill defined, or is fuzzy \cite{madore}.
Indeed in M-Theory too, we have a noncommutative geometry like
((\ref{e3z})). As we started with a minimum extention at the Compton
scale, let us
take $a = (l, \tau)$.\\
Then the above conclusion is in fact true, because as discussed in
detail \cite{cu,weinberg}, by virtue of the Heisenberg Uncertainty
Principle, there are unphysical superluminal effects within this
scale.\\
Another way of seeing this is by starting from the usual Dirac
coordinate \cite{pdirac}
\begin{equation}
x_\imath = \left(c^2 p_\imath H^{-1}t\right) + \frac{1}{2} c\hbar \left(\alpha_\imath - cp_\imath H^{-1}\right) H^{-1}\label{e9}
\end{equation}
where the $\alpha$'s are given by
\begin{equation}
\vec \alpha = \left[\begin{array}{ll}
\vec \sigma \quad 0\\
0 \quad \vec \sigma
\end{array}
\right]\quad \quad ,\label{e10}
\end{equation}
the $\sigma$'s being the usual Pauli matrices. The first term on the right side of (\ref{e9}) is the usual Hermitian position coordinate. It is the second or imaginary term which contains $\vec{\alpha}$ that makes the Dirac coordinate non Hermitian. However we can easily verify from the commutation relations of $\vec{\alpha}$, using (\ref{e10}) that
\begin{equation}
[x_\imath , x_j] = \beta_{\imath j} \cdot l^2\label{eA}
\end{equation}
In fact (\ref{eA}) is just a form of the first of equations (\ref{e3z}) and brings out the fuzzyness of spacetime in intervals where order of $l^2$ is not neglected.\\
Dirac himself noticed this feature of his coordinate and argued \cite{pdirac} that our physical spacetime is actually one in which averages at the Compton scale are taken. Effectively he realized that point spacetime is not physical. Once such averages are taken, he pointed out that the rapidly oscillating second term in (\ref{e9}) or zitterbewegung gets eliminated.\\
We now obtain a rationale for the Dirac equation and spin from (\ref{eA}) \cite{bgscsf,bgsnc}. Under a time elapse transformation of the wave
function, (or, alternatively, as a small scale transformation),
\begin{equation}
| \psi' > = U(R)| \psi >\label{e8a}
\end{equation}
we get
\begin{equation}
\psi' (x_j) = [1 + \imath \epsilon (\imath x_j \frac{\partial}{\partial x_j}) + 0 (\epsilon^2)] \psi
(x_j)\label{e9a}
\end{equation}
Equation (\ref{e9a}) has been shown to lead to the Dirac equation
when $\epsilon$ is the Compton time. A quick way to see this is as follows: At the Compton scale we have,
$$|\vec {L} | = | \vec {r} \times \vec {p} | = | \frac{\hbar}{2mc} \cdot mc| = \frac{\hbar}{2},$$
that is, at the Compton scales we get the Quantum Mechanical spin
from the usual angular momentum. Next, we can easily verify, that
the choice,
\begin{equation}
t = \left(\begin{array}{ll}
1 \quad 0\\
0 \quad -1
\end{array}
\right), \vec {x} = \left(\begin{array}{ll}
0 \quad \vec {\sigma}\\
\vec {\sigma} \quad 0
\end{array}
\right)\label{ed}
\end{equation}
provides a representation for the coordinates in (\ref{e3z}), apart
from scalar factors. As can be seen, this is also a representation
of the Dirac matrices. Substitution of the above in (\ref{e9a})
leads to the Dirac equation
$$(\gamma^\mu p_\mu - mc^2) \psi = 0$$
because
$$E\psi = \frac{1}{\epsilon}\{\psi' (x_j) - \psi (x_j)\}, \quad E = mc^2,$$
where $\epsilon = \tau$ (Cf.ref.\cite{wolf}).\\
All this is symptomatic of an underlying fuzzy spacetime described
by a noncommutative space time geometry
(\ref{eA}) or (\ref{e3z}) \cite{sakharov}.\\
The point here is that under equation (\ref{eA}) and (\ref{ed}), the coordinates
$x^\mu \to \gamma^{(\mu)} x^{(\mu)}$ where the brackets with the
superscript denote the fact that there is no summation over the
indices.  Infact, in the theory of the Dirac equation it is well
known \cite{bade}that,
\begin{equation}
\gamma^k \gamma^l + \gamma^l \gamma^k = - 2g^{kl}I\label{e10a}
\end{equation}
where $\gamma$'s satisfy the usual Clifford algebra of the Dirac
matrices, and can be represented by
\begin{equation}
\gamma^k = \sqrt{2} \left(\begin{array}{ll}
0 \quad \sigma^k \\
\sigma^{k*} \quad 0
\end{array}\right)\label{e11a}
\end{equation}
where $\sigma$'s are the Pauli matrices. Bade and Jehle noted that
(Cf.ref.\cite{bade}), we could take the $\sigma$'s or $\gamma$'s in
(\ref{e11a}) and (\ref{e10a}) as the components of a contravariant
world vector, or equivalently we could take them to be fixed
matrices, and to maintain covariance, to attribute new
transformation properties to the wave function, which now becomes a
spinor (or bi-spinor). This latter has been the traditional route,
because of which the Dirac wave function has its bi-spinorial
character. In this latter case, the coordinates retain their usual
commutative or point character. It is only when we consider the
equivalent former alternative, that we return to the noncommutative
geometry (\ref{eA}).\\
That is, in the usual commutative spacetime the Dirac spinorial
wave functions conceal the noncommutative
character (\ref{eA}).
\section{Branes}
The considerations leading from (\ref{eA}) to (\ref{e11a}) show that
we are essentially dealing with a Clifford or C-space \cite{pavsic}.
We will study this briefly, following \cite{pavsic}. Given the
$\gamma$ matrices which we encountered earlier we can write
\begin{equation}
\gamma_\mu \cdot \gamma_\nu \equiv \frac{1}{2} (\gamma_\mu
\gamma_\nu + \gamma_\nu \gamma_\mu ) = g_{\mu \nu}.\label{e17}
\end{equation}
\begin{equation}
\gamma_\mu \wedge \gamma_\nu = \frac{1}{2} (\gamma_\mu \gamma_\nu -
\gamma_\nu \gamma_\mu ) \equiv \frac{1}{2} [\gamma_\mu , \gamma_\nu
].\label{e18}
\end{equation}
In other words (\ref{e18}) gives an antisymmetrical tensor.\\
More generally we can consider a complete set of basis vectors
$\gamma_\mu$ in a n-dimensional space satisfying (\ref{e17}) and
(\ref{e18}). We can then have
\begin{equation}
\gamma_{\mu_1} \wedge \gamma_{\mu_2} \wedge \gamma_{\mu_3} =
\frac{1}{3!} [\gamma_{\mu_1},
\gamma_{\mu_2},\gamma_{\mu_3}],\label{e19}
\end{equation}
\begin{equation}
\vdots\label{e20}
\end{equation}
\begin{equation}
\gamma_{\mu_1} \wedge \gamma_{\mu_2} \wedge \cdots \wedge
\gamma_{\mu_n} = \frac{1}{r!} [\gamma_{\mu_1}, \gamma_{\mu_2},\cdots
, \gamma_{\mu_n}].\label{e21}
\end{equation}
The left sides of (\ref{e19}), (\ref{e20}) and (\ref{e21}) are
termed $p$-vectors, where $p$ takes on the values, $3,4,\cdots n$. A
point in this n-dimensional space can be designated as in the
previous section by
\begin{equation}
x = x^\mu \gamma_\mu .\label{e22}
\end{equation}
More generally we have poly vectors which are obtained by
superposing multivectors as follows
\begin{equation}
X = \sigma 1 + x^\mu \gamma_\mu + \frac{1}{2} x^{\mu_1 \mu_2}
\gamma_{\mu_1 \mu_1} + \cdots + \frac{1}{n!} x^{\mu_1 \cdots \mu_n}
\gamma_{\mu_1 \cdots \mu_n} \equiv x^M \gamma_M.\label{e23}
\end{equation}
where
$$\gamma_{\mu_1 \cdots \mu_r} \equiv \gamma_{\mu_1} \wedge
\gamma_{\mu_2} \wedge \cdots \wedge \gamma_{\mu_r}$$ and
$$x^M = (\sigma , x^\mu , x^{\mu_1\mu_2},\cdots ,x^{\mu_1 \cdots
\mu_r},$$
\begin{equation}
\gamma_M = (1, \gamma_\mu , \gamma_{\mu_1\mu_2,\cdots},
\gamma_{\mu_1 \cdots \mu_r}), \quad \mu_1 < \mu_2 < \cdots <
\mu_r\label{e24}
\end{equation}
The coordinates $X_{\mu_1 \cdots \mu_p}$ is a $p$-area enclosed by a
loop of dimension $p-1$. We now observe that the coordinates
$\sigma, x_\mu , x_{\mu_1\mu_2}$ etc. describe extended objects and
that $x^M$ is a quantity that assumes any real value and that all
possible $X$ forms constitute a $2^n$ dimensional manifold, which we
call the C-space. It may be mentioned that such higher dimensional
extended objects or surfaces- the D-branes, were introduced by Polchinski \cite{pol}.\\
In any case we can see that the C-space generates branes of
different dimensionality as in M-Theory. If we stop with $x_\mu$, we
have the point space time of bosons, if we stop with $x^\mu
\gamma_\mu$ (in (\ref{e23})), then we have the Fermions (of the
earlier section) and finally we get branes by retaining other terms
in (\ref{e23}). Moreover as we saw, retaining the usual coordinates
$x^\mu$ tantamounts to neglecting $O(l^2)$, while for Fermions we
retain those terms which are $\sim 10^{-22}cm$ in our Compton
wavelength description, while if we retain term $O(l^3)$ for
example, these are $\sim 10^{-33}cm$ (the Planck scale) and so on.
However, we are really dealing with areas, volumes etc. in these
higher order terms, and fractal dimensions as these are resolution
dependent.

\end{document}